\documentclass[eqsecnum,superscriptaddress,showpacs]{revtex4}
\usepackage[dvips]{graphicx}
\usepackage{latexsym}
\usepackage{bm}
\newcommand{\Slash}[1]{{\ooalign{\hfil/\hfil\crcr$#1$}}}

\begin{document}

\title{Light-Light Scattering}

\author{Naohiro Kanda}\email{nkanda@phys.cst.nihon-u.ac.jp}
\affiliation{Department of Physics, Faculty of Science and Technology, 
Nihon University, Tokyo, Japan}

\date{\today}%

\begin{abstract}


For a long time, it is believed that the light by light scattering is described 
properly by the Lagrangian density obtained by Heisenberg and Euler. Here, we 
present a new calculation which is based on the modern field theory technique. 
It is found that the light-light scattering is completely different from the old 
expression. The reason is basically due to the unphysical condition (gauge condition)
which was employed by the QED calcualtion of Karplus and Neumann.  
The correct cross section of light-light scattering at low energy of 
$(\frac{\omega }{m} \ll 1)$ can be written as
$ \displaystyle{ \frac{d\sigma }{d\Omega }=\frac{1}{(6\pi )^2}\frac{\alpha ^4}
{(2\omega )^2}(3+2\cos^2\theta +\cos^4\theta)  }$.

\end{abstract}


\maketitle


\section{Introduction}

It has been believed that photon-photon scattering can be 
described by the Lagrangian density of Heisenberg and Euler \cite{EK}\cite{Euler}\cite{HE}. 
In addition, this cross section is confirmed by the QED calculation of 
Karplus and Neumann \cite{LL}\cite{KN}\cite{KN2}. However, the calculation by 
Karplus and Neumann is not reliable since they have put some additional conditions 
on the QED calculation so as to reproduce the result obtained by Heisenberg and Euler. 

In this respect, it is important that the proper treatment of the photon-photon 
scattering should be made again at the present stage. 
The calculation of photon-photon scattering itself is straightforward since 
the Feynman diagrams of the fourth order perturbation calculation can be done 
without any difficulties. 
The interaction Lagrangian density of $ H'=-ej_\mu A^\mu $ is gauge invariant 
as long as the fermion current is conserved, which is indeed the case. 
Therefore, the QED calculation itself has no conceptual difficulty and therefore 
we should carry out the S-matrix evaluation of photon-photon scattering. 

\begin{figure}[htbp]
\begin{center}
 \begin{minipage}{0.30\hsize}
  \begin{center}
   \includegraphics[width=40mm]{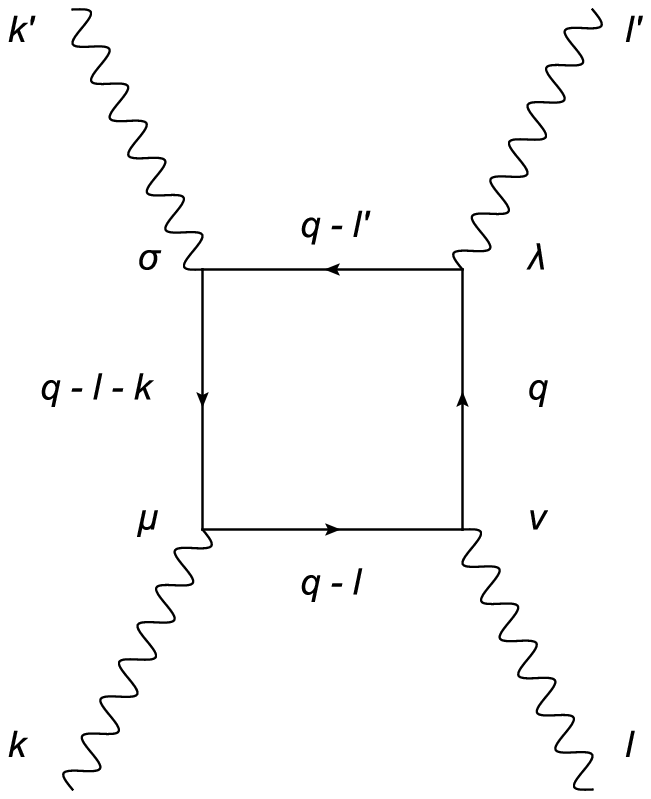}
  \end{center}
  \caption{$M_a$}
  \label{fig:one}
 \end{minipage}
 \begin{minipage}{0.30\hsize}
 \begin{center}
  \includegraphics[width=34mm]{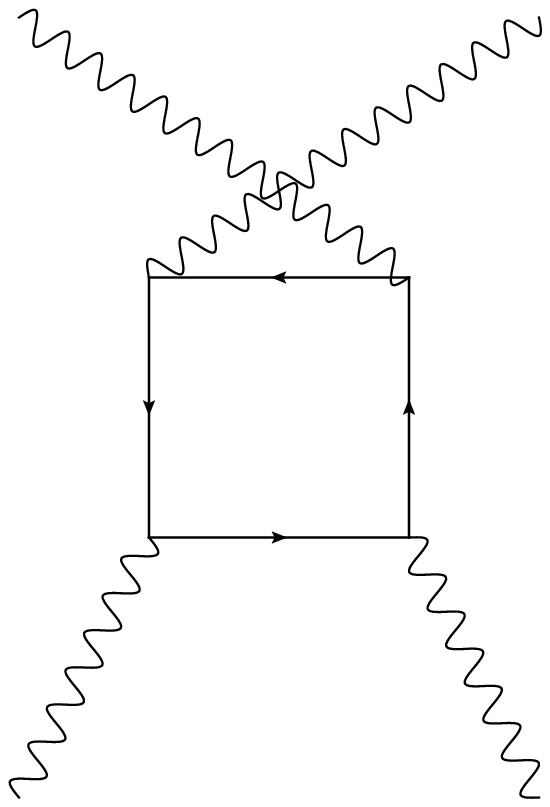}
 \end{center}
  \caption{$M_b$}
  \label{fig:two}
 \end{minipage}
 \begin{minipage}{0.30\hsize}
 \begin{center}
  \includegraphics[width=34mm]{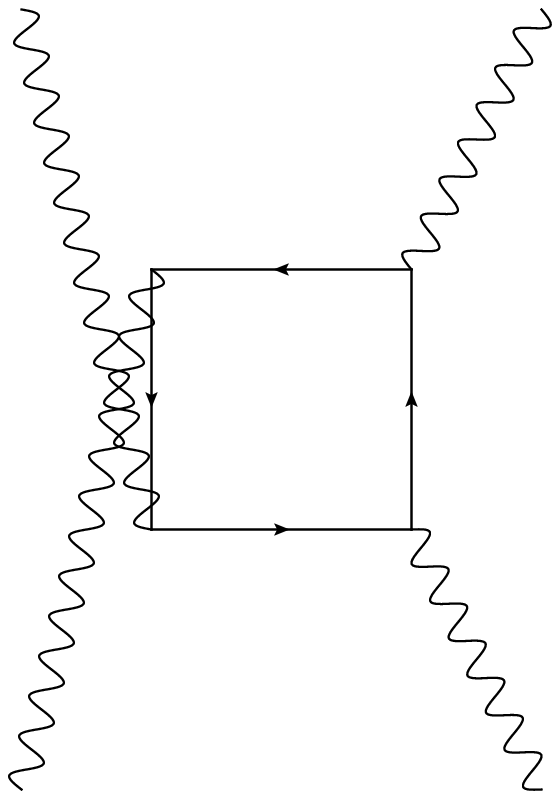}
 \end{center}
  \caption{$M_c$}
  \label{fig:three}
 \end{minipage}
\end{center} 
\end{figure}

\newpage

\section{Feynman Amplitude of Photon-Photon Scattering}
The Feynman amplitude of the photon-photon scattering can be written as
$$ \hspace*{-6.3cm} M^{rr'ss'}=2(M_a+M_b+M_c) \eqno{(1.1)} $$ 
$$ =2(\bar{M}^{\mu \nu \lambda \sigma }_a+\bar{M}^{\mu \nu \lambda \sigma }_b
+\bar{M}^{\mu \nu \lambda \sigma }_c)
\epsilon ^{r}_{\mu }(k)\epsilon ^{s}_{\nu }(l)
\epsilon ^{s'}_{\lambda }(l')\epsilon ^{r'}_{\sigma }(k') $$
where we note that the Feynman amplitude is not affected by the direction 
of the loop momentum  \cite{nishi}. 
The amplitude $ M_a$ can be explicitly written as
$$ M_a = -\frac{(ie)^4}{(2\pi)^4} \int d^4q 
\frac{ \text{Tr} [\gamma ^{\mu }(\Slash{q}-\Slash{l}-\Slash{k}+m)
\gamma ^{\sigma }(\Slash{q}-\Slash{l}'+m)\gamma ^{\lambda }(\Slash{q}+m)
\gamma ^{\nu }(\Slash{q}-\Slash{l}+m)]}
{[(q-l-k)^2-m^2+i\varepsilon ][(q-l')^2-m^2+i\varepsilon ]
[q^2-m^2+i\varepsilon ][(q-l)^2-m^2+i\varepsilon ]} $$
$$ \hspace*{10cm} \times \epsilon ^{r}_{\mu }(k)\epsilon ^{r'}_{\sigma }(k')
\epsilon ^{s'}_{\lambda }(l')\epsilon ^{s}_{\nu }(l) . \eqno{(1.2)} $$
Here, it should be important to note that the total Feynman amplitude of the 
photon-photon scattering does not have any logarithmic divergence even though 
the amplitude $M_a$ alone has the logarithmic divergence. 
This is quite important in that the physical processes should not have any 
divergences, and indeed the photon-photon scattering is just the case. 
This strongly suggests that the evaluation of the Feynman amplitude of 
the photon-photon scattering process must be directly connected to 
the real physical process which should be observed by experiments. 

\section{Divergence}
The amplitude $M_a$ has the logarithmic divergence, and this can be seen when 
we check the large $q$ behavior \cite{LL}. In this case, we find from (1.2)
$$ M_a \sim -\frac{(ie)^4}{(2\pi)^4} \int d^4q \frac{4}{3} \frac{1}{(q^2)^2}
\{g^{\mu \nu }g^{\lambda \sigma }+g^{\mu \sigma }g^{\nu \lambda }
-2g^{\nu \sigma }g^{\mu \lambda } \} 
\epsilon ^{r}_{\mu }\epsilon ^{r}_{\nu }\epsilon ^{s'}_{\lambda }
\epsilon ^{r'}_{\sigma } \eqno{(2.1)} $$
which has obviously the logarithmic divergnce. However, if we add all of the 
amplitude together, then we find from (1.1)
$$ \hspace*{-4.2cm} M \sim  -2 \frac{(ie)^4}{(2\pi)^4} \frac{4}{3} 
\int d^4q \frac{1}{(q^2)^2}\{g^{\mu \nu }g^{\lambda \sigma }
+g^{\mu \sigma }g^{\nu \lambda }-2g^{\nu \sigma }g^{\mu \lambda }
$$
$$
+g^{\mu \nu }g^{\sigma \lambda }+g^{\mu \lambda }g^{\nu \sigma }-2g^{\nu \lambda }g^{\mu \sigma }
\eqno{(2.2)} $$
$$ \hspace*{1.5cm}
+g^{\sigma \nu }g^{\lambda \mu }+g^{\sigma \mu }g^{\nu \lambda }-2g^{\nu \mu }g^{\sigma \lambda }
\} 
\epsilon ^{r}_{\mu }\epsilon ^{r}_{\nu }\epsilon ^{s'}_{\lambda }\epsilon ^{r'}_{\sigma }
$$
$$ \hspace*{-11cm} = 0 . $$
This means that the total amplitude has no divergence at all because of the cancellation, 
and it is indeed finite. Therefore, we do not have to employ
 any specific regularization scheme, and  thus the evaluation is very reliable.

\section{Definition of Polarization vector}
Here, we take the polarization vector as defined by Lifshitz \cite{LL}
$$ \bm{\epsilon}^{(1)}_{1} = \bm{\epsilon}^{(1)}_{2}
 = \bm{\epsilon}^{(1)}_{3} = \bm{\epsilon}^{(1)}_{4}
=\frac{\bm{k} \times \bm{k}'}{|\bm{k} \times \bm{k}'|} \hspace*{2cm}
\bm{\epsilon}^{(2)}_{1} = \frac{1}{\omega} \bigl( \bm{k} \times \bm{\epsilon}^{(1)}_{1} \bigr)
=-\bm{\epsilon}^{(2)}_{2} $$
$$ \hspace*{-7.3cm}
 \bm{\epsilon}^{(2)}_{4} = \frac{1}{\omega} \bigl( \bm{k}' \times 
\bm{\epsilon}^{(1)}_{4} \bigr) = -\bm{\epsilon}^{(2)}_{3} $$
$ \bm{e}^{(i)}_{n} $ denotes the polarization vector of photons. 
Here, we take the Coulomb gauge fixing $ \bm{\nabla} \cdot \bm{A} = 0 $. 

Each photon has the following momenta 
\vspace*{0.5\baselineskip}\\
\hspace*{9.62216pt}\text{initial state}
$$ \text{photon \ 1} \ : \ k^{\mu} = (\omega,\bm{k}) \hspace*{2cm}
 \text{photon \ 2} \ : \ l^{\mu} = (\omega,\bm{l}) $$
\vspace*{0.5\baselineskip}\\
\hspace*{9.62216pt}\text{final state}
$$ \text{photon \ 3} \ : \ l'^{\mu} = (\omega,\bm{l}') \hspace*{2cm}
 \text{photon \ 4} \ : \ k'^{\mu} = (\omega,\bm{k}') . $$
$$ \omega = |\bm{k}| = |\bm{l}| = |\bm{k}'| = |\bm{l}'| $$
Also, by noting that there is no rest system, we find
$$ \bm{l} = -\bm{k} \hspace*{2cm} \bm{l}' = -\bm{k}' . $$

\section{Calculation of $M_a$ at Low Energy}
Now, we carry out the calculation
of $M_a$ as an example
$$ M_a = \bar{M}^{\mu \nu \lambda \sigma }_a
\epsilon ^{r}_{\mu }(k)\epsilon ^{s}_{\nu }(l)
\epsilon ^{s'}_{\lambda }(l')\epsilon ^{r'}_{\sigma }(k') \eqno{(4.1)} $$
where $\bar{M}^{\mu \nu \lambda \sigma}_a$ can be written as
$$ \bar{M}_a^{\mu \nu \lambda \sigma } = -\frac{(ie)^4}{(2\pi)^4} \int d^4q 
\frac{ \text{Tr} [\gamma ^{\mu }(\Slash{q}-\Slash{l}-\Slash{k}+m)
\gamma ^{\sigma }(\Slash{q}-\Slash{l}'+m)\gamma ^{\lambda }(\Slash{q}+m)
\gamma ^{\nu }(\Slash{q}-\Slash{l}+m)]}
{[(q-l-k)^2-m^2+i\varepsilon ][(q-l')^2-m^2+i\varepsilon ]
[q^2-m^2+i\varepsilon ][(q-l)^2-m^2+i\varepsilon ]}. \eqno{(4.2)} $$
By making use of the  Feynman parameter \cite{mandl}, we find 
$$ \hspace*{.8cm} \bar{M}_a^{\mu \nu \lambda \sigma }
 = -\frac{(ie)^4}{(2\pi)^4} \int d^4q
 \ 3! \int^{1}_{0}dz_{1} \int^{z_1}_{0}dz_{2} \int^{z_2}_{0}dz_{3}
\frac{ \text{Tr} [\gamma ^{\mu }(\Slash{q}-\Slash{l}-\Slash{k}+m)
\gamma ^{\sigma }(\Slash{q}-\Slash{l}'+m)\gamma ^{\lambda }(\Slash{q}+m)
\gamma ^{\nu }(\Slash{q}-\Slash{l}+m)]}
{[q^2-2q . A+B+i\varepsilon ]^4} $$
where
$$ A^{\mu} \equiv l^{\mu}z_{1}+(l'-l)^{\mu}z_{2}+k'^{\mu}z_{3} $$
$$ \hspace*{-1.8cm} B \equiv  2l . kz_{3}-m^2 $$
By putting $q-A=t$, we find 
$$ \hspace*{-10cm}
\bar{M}_a^{\mu \nu \lambda \sigma } = -\frac{(ie)^4}{(2\pi)^4}
3! \int^{1}_{0}dz_{1} \int^{z_1}_{0}dz_{2} \int^{z_2}_{0}dz_{3} $$
$$ \hspace*{2cm} \times \int d^4t 
\frac{ \text{Tr} [\gamma ^{\mu }(\Slash{t}+\Slash{A}-\Slash{l}-\Slash{k}+m)
\gamma ^{\sigma }(\Slash{t}+\Slash{A}-\Slash{l}'+m)\gamma ^{\lambda }
(\Slash{t}+\Slash{A}+m)\gamma ^{\nu }(\Slash{t}+\Slash{A}-\Slash{l}+m)]}
{[t^2-A^2+B+i\varepsilon ]^4}. $$
Here, for simplicity, we define
$$ \Slash{a} = \Slash{A}-\Slash{l}-\Slash{k} $$
$$ \hspace*{-.6cm} \Slash{b} = \Slash{A}-\Slash{l}' $$
$$ \hspace*{-1.2cm} \Slash{c} = \Slash{A} $$
$$ \hspace*{-.6cm} \Slash{d} = \Slash{A}-\Slash{l} $$
\newpage
and thus 
$$ \bar{M}_a^{\mu \nu \lambda \sigma } = -\frac{(ie)^4}{(2\pi)^4}
3! \int^{1}_{0}dz_{1} \int^{z_1}_{0}dz_{2} \int^{z_2}_{0}dz_{3} \int d^4t 
\frac{ \text{Tr} [\gamma ^{\mu }(\Slash{t}+\Slash{a}+m)
\gamma ^{\sigma }(\Slash{t}+\Slash{b}+m)\gamma ^{\lambda }(\Slash{t}+\Slash{c}+m)
\gamma ^{\nu }(\Slash{t}+\Slash{d}+m)]}
{[t^2-A^2+B+i\varepsilon ]^4} $$
In the numerator, we find
 $ \text{Tr} [ \text{odd-numbers of} \ \gamma \text{-matrices} ]=0$ and by noting
$\int d^4t \ \underbrace{t^{\mu }t^{\nu }\cdots t^{\lambda }}_{odd} \ f(t^2)=0$
we find
$$ \hspace*{-7cm} \text{Tr} [\gamma ^{\mu }(\Slash{t}+\Slash{a}+m)
\gamma ^{\sigma }(\Slash{t}+\Slash{b}+m)\gamma ^{\lambda }(\Slash{t}+\Slash{c}+m)
\gamma ^{\nu }(\Slash{t}+\Slash{d}+m)] $$
$$ = t^2 F^{\mu \nu \lambda \sigma}+G^{\mu \nu \lambda \sigma}
+ \text{Tr} [\gamma ^{\mu }\Slash{t}\gamma ^{\sigma }\Slash{t}
\gamma ^{\lambda }\Slash{t}\gamma ^{\nu }\Slash{t}]  $$
where
$$ \hspace*{-4cm} F^{\mu \nu \lambda \sigma} \equiv
 -4m^2 \{g^{\mu \sigma }g^{\lambda \nu }+g^{\mu \nu }g^{\sigma \lambda }
-2g^{\mu \lambda }g^{\sigma \nu } \} $$
$$ -\frac{1}{2} \text{Tr} [ \{
\gamma ^{\mu }\gamma ^{\sigma }\gamma ^{\lambda }\Slash{c}\gamma ^{\nu }\Slash{d}
+\gamma ^{\mu }\gamma ^{\lambda }\Slash{b}\gamma ^{\sigma }\gamma ^{\nu }\Slash{d}
+\gamma ^{\mu }\gamma ^{\sigma }\Slash{b}\gamma ^{\lambda }\Slash{c}\gamma ^{\nu } $$
$$ +(\gamma ^{\mu }\Slash{a}\gamma ^{\sigma })(\gamma ^{\lambda }\gamma ^{\nu }\Slash{d}
+\gamma ^{\nu }\Slash{c}\gamma ^{\lambda }+\Slash{b}\gamma ^{\lambda }\gamma ^{\nu }) \} ] $$
$$ \hspace*{-2.5cm} G^{\mu \nu \lambda \sigma} \equiv
 \text{Tr} [\gamma ^{\mu }(\Slash{a}+m)\gamma ^{\sigma }(\Slash{b}+m)
\gamma ^{\lambda }(\Slash{c}+m)\gamma ^{\nu }(\Slash{d}+m)] . $$ 
Now, we find
$$ \bar{M}^{\mu \nu \lambda \sigma}_a = -\frac{(ie)^4}{(2\pi)^4}
3! \int^{1}_{0}dz_{1} \int^{z_1}_{0}dz_{2} \int^{z_2}_{0}dz_{3} \int d^4t 
\frac{t^2 F^{\mu \nu \lambda \sigma}+G^{\mu \nu \lambda \sigma}
+ \text{Tr} [\gamma ^{\mu }\Slash{t}\gamma ^{\sigma }\Slash{t}
\gamma ^{\lambda }\Slash{t}\gamma ^{\nu }\Slash{t}]}
{(t^2-A^2+B+i\varepsilon )^4} $$
and we carry out the integration and obtain \cite{mandl}
$$ \hspace*{-3.2cm}
 \bar{M}^{\mu \nu \lambda \sigma}_a = -\frac{(ie)^4}{(2\pi)^4}
\int^{1}_{0}dz_{1} \int^{z_1}_{0}dz_{2} \int^{z_2}_{0}dz_{3}
i\pi^2 \biggl\{ \frac{2F^{\mu \nu \lambda \sigma}}{B-A^2}
+\frac{G^{\mu \nu \lambda \sigma}}{(B-A^2)^2} $$
$$ \hspace*{6cm} +8(g^{\mu \sigma }g^{\lambda \nu }+g^{\mu \nu }g^{\lambda \sigma }
-2g^{\sigma \nu }g^{\mu \lambda })
\biggr(\ln\biggl|\frac{\Lambda ^2}{B-A^2}\biggr|-\frac{11}{6}\biggr) \biggr\} $$
where $\Lambda $ denotes the cutoff momentum. 

Here, we consider the case of $\frac{\omega }{m} \ll 1$ and 
expand the integration in terms of $\frac{\omega}{m}$ powers. 
This approximation is quite reasonable as we show it later \cite{FK}.


Now, we consider the expansion in terms of $\frac{\omega}{m}$. Noting that the $m$ 
never appears in $k, \ \ l, \ \ k', \ \ l' $ in $F^{\mu \nu \lambda \sigma},
\ \ G^{\mu \nu \lambda \sigma} $ but only $\omega$ appears, we can rewrite 
$F^{\mu \nu \lambda \sigma}, \ \ G^{\mu \nu \lambda \sigma} $ and obtain
$$ \hspace*{-1.4cm} F^{\mu \nu \lambda \sigma}
 = -4m^2Q^{\mu \nu \lambda \sigma}-\frac{1}{2}\omega^2 R^{\mu \nu \lambda \sigma} $$ 
$$ G^{\mu \nu \lambda \sigma}
 = \omega^4S^{\mu \nu \lambda \sigma}+m^2 \omega^2 R^{\mu \nu \lambda \sigma}
+m^4 T^{\mu \nu \lambda \sigma} $$
where
$$ \hspace*{-3.4cm} Q^{\mu \nu \lambda \sigma }
 \equiv  g^{\mu \sigma }g^{\lambda \nu }
+g^{\mu \nu }g^{\sigma \lambda }-2g^{\mu \lambda }g^{\sigma \nu } $$
$$ R^{\mu \nu \lambda \sigma } \equiv \frac{1}{\omega^2} \text{Tr} [ \{
\gamma ^{\mu }\gamma ^{\sigma }\gamma ^{\lambda }\Slash{c}\gamma ^{\nu }\Slash{d}
+\gamma ^{\mu }\gamma ^{\lambda }\Slash{b}\gamma ^{\sigma }\gamma ^{\nu }\Slash{d}
+\gamma ^{\mu }\gamma ^{\sigma }\Slash{b}\gamma ^{\lambda }\Slash{c}\gamma ^{\nu } $$
$$ +(\gamma ^{\mu }\Slash{a}\gamma ^{\sigma })(\gamma ^{\lambda }\gamma ^{\nu }\Slash{d}
+\gamma ^{\nu }\Slash{c}\gamma ^{\lambda }+\Slash{b}\gamma ^{\lambda }\gamma ^{\nu }) \} ] $$
$$ \hspace*{-4.2cm} S^{\mu \nu \lambda \sigma }
 \equiv \frac{1}{\omega^4} \text{Tr} [\gamma ^{\mu }\Slash{a}\gamma ^{\sigma }\Slash{b}
\gamma ^{\lambda }\Slash{c}\gamma ^{\nu }\Slash{d}] $$
$$ \hspace*{-5.4cm} T^{\mu \nu \lambda \sigma }
 \equiv \text{Tr} [\gamma ^{\mu }\gamma ^{\sigma }
\gamma ^{\lambda }\gamma ^{\nu }] $$
Here, $ Q^{\mu \nu \lambda \sigma}, \ \ R^{\mu \nu \lambda \sigma}, 
\ \ S^{\mu \nu \lambda \sigma}, \ \ T^{\mu \nu \lambda \sigma} $ 
are all dimensionless. In this case, we can easily expand 
$ Q^{\mu \nu \lambda \sigma}, \ \ R^{\mu \nu \lambda \sigma}, \
 \ S^{\mu \nu \lambda \sigma}, \ \ T^{\mu \nu \lambda \sigma} $.

\newpage
Now, we can write 
$$ B-A^2 = -m^2 \biggl\{ 1-\frac{\omega^2}{m^2}U_a \biggr\} $$
$$ U_a = 4\Bigl[z_{2}(z_{2}-z_{1}-z_{3})\sin^2\frac{\theta }{2}
-z_{3}z_{1}\cos^2\frac{\theta }{2}+z_{3} \Bigr] $$
and therefore we have
$$ \bar{M}^{\mu \nu \lambda \sigma}_a = -\frac{(ie)^4}{(2\pi)^4} i\pi^2
\int^{1}_{0}dz_{1} \int^{z_1}_{0}dz_{2} \int^{z_2}_{0}dz_{3}
\biggl\{ \frac{2F^{\mu \nu \lambda \sigma }}{B-A^2}
+\frac{G^{\mu \nu \lambda \sigma }}{(B-A^2)^2}
+8Q^{\mu \nu \lambda \sigma }
\biggr(\ln\biggl|\frac{\Lambda ^2}{m^2}\biggr|-\ln|1-\xi ^2 U_a|
-\frac{11}{6}\biggr) \biggr\} $$
$$ \hspace*{-10cm} \xi \equiv \frac{\omega}{m}. $$
Here, we omit the index of $ \mu, \  \nu, \  \lambda, \  \sigma $ in 
$ Q^{\mu \nu \lambda \sigma}, \  R^{\mu \nu \lambda \sigma}, \
 S^{\mu \nu \lambda \sigma}, \ T^{\mu \nu \lambda \sigma} $. 
Therefore, we find
$$ \hspace*{-4cm}
\bar{M}_a^{\mu \nu \lambda \sigma } = -\frac{(ie)^4}{(2\pi)^4} i\pi^2
\int^{1}_{0}dz_{1} \int^{z_1}_{0}dz_{2} \int^{z_2}_{0}dz_{3}
\biggl\{ \frac{-8m^2 Q-\omega^2 R}{-m^2(1-\xi^2 U)}
+\frac{\omega^4 S+m^2 \omega^2 R+m^4 T}{(-m^2(1-\xi^2 U))^2} $$
$$ \hspace*{10cm} +8Q
\biggr(\ln\biggl|\frac{\Lambda ^2}{m^2}\biggr|-\ln|1-\xi ^2 U|
-\frac{11}{6}\biggr) \biggr\} $$
$$ \hspace*{-5.6cm}
\bar{M}_a^{\mu \nu \lambda \sigma } = -\frac{(ie)^4}{(2\pi)^4} i\pi^2
\int^{1}_{0}dz_{1} \int^{z_1}_{0}dz_{2} \int^{z_2}_{0}dz_{3}
\biggl\{ \frac{8Q +\xi^2 R}{1-\xi^2 U}
+\frac{\xi^4 S+\xi^2 R+T}{(1-\xi^2 U)^2} $$
$$ \hspace*{10cm} +8Q
\biggr(\ln\biggl|\frac{\Lambda ^2}{m^2}\biggr|-\ln|1-\xi ^2 U|
-\frac{11}{6}\biggr) \biggr\} $$
Expanding in terms of $\xi = \frac{\omega}{m}$, we find
$$ \hspace*{-3.4cm}
\bar{M}_a^{\mu \nu \lambda \sigma } = -\frac{(ie)^4}{(2\pi)^4} i\pi^2
\int^{1}_{0}dz_{1} \int^{z_1}_{0}dz_{2} \int^{z_2}_{0}dz_{3}
\biggl\{ 8Q+(R+8QU)\xi^2+(RU+8QU^2)\xi^4 +\cdots $$
$$ \hspace*{4.1cm} +T+(R+2TU)\xi^2+(S+2RU+3TU^2)\xi^4 +\cdots $$
$$ \hspace*{3.4cm} +8Q
\biggr(\ln\biggl|\frac{\Lambda ^2}{m^2}\biggr|+U\xi ^2 +\frac{1}{2}\xi^4 U^2+\cdots
-\frac{11}{6}\biggr) \biggr\} $$
$$ \hspace*{-4.6cm}
\bar{M}_a^{\mu \nu \lambda \sigma } = -\frac{(ie)^4}{(2\pi)^4} i\pi^2
\int^{1}_{0}dz_{1} \int^{z_1}_{0}dz_{2} \int^{z_2}_{0}dz_{3}
\biggl\{ 8Q
+8Q \biggl( \ln \biggl| \frac{\Lambda ^2}{m^2} \biggr|
-\frac{11}{6} \biggr)+T+\cdots \biggr\} $$


The coefficient of Q at the lowest order can be
 written as $ 8 + 8 \bigr( \ln \bigr| \frac{\Lambda^2}{m^2} \bigr| -\frac{11}{6} \bigl) $,
 which does not depend on the shape of $ M_{a}, \ M_{b}, \ M_{c} $.
Therefore, we find
$$ \hspace*{-7.2cm}
\bar{M}_a^{\mu \nu \lambda \sigma } \cong -\frac{(ie)^4}{(2\pi)^4} i\pi^2
\int^{1}_{0}dz_{1} \int^{z_1}_{0}dz_{2} \int^{z_2}_{0}dz_{3} \ T
 = -\frac{(ie)^4}{(2\pi)^4} i\pi^2 \frac{1}{6} \ T $$
and we can write it explicitly as 
$$ \bar{M}_a^{\mu \nu \lambda \sigma }
 = -\frac{(ie)^4}{(2\pi)^4} i\pi^2 \frac{1}{6} \ T^{\mu \nu \lambda \sigma } . $$

\section{Total Amplitude}
In this way, we can obtain the shape of $\bar{M}^{\mu \nu \lambda \sigma}_a$, \
$\bar{M}^{\mu \nu \lambda \sigma}_b$ and $\bar{M}^{\mu \nu \lambda \sigma}_c$ 
at $\frac{\omega}{m} \ll 1$. 
Thus, we can write for $M_a, \  M_b, \  M_c$ as
$$ M_a = -\frac{(ie)^4}{(2\pi)^4} i\pi^2 \frac{1}{6}
\biggl\{ 4(g^{\mu \sigma }g^{\lambda \nu }+g^{\mu \nu }g^{\sigma \lambda }
-g^{\mu \lambda }g^{\sigma \nu }) \biggr\}
\epsilon ^{r}_{\mu }\epsilon ^{s}_{\nu }\epsilon ^{s'}_{\lambda }\epsilon ^{r'}_{\sigma }
 \eqno{(5.1.a)} $$

$$ M_b = -\frac{(ie)^4}{(2\pi)^4} i\pi^2 \frac{1}{6}
\biggl\{ 4(g^{\mu \lambda }g^{\sigma \nu }+g^{\mu \nu }g^{\sigma \lambda }
-g^{\mu \sigma }g^{\lambda \nu }) \biggr\}
\epsilon ^{r}_{\mu }\epsilon ^{s}_{\nu }\epsilon ^{s'}_{\lambda }\epsilon ^{r'}_{\sigma }
 \eqno{(5.1.b)} $$

$$ M_c = -\frac{(ie)^4}{(2\pi)^4} i\pi^2 \frac{1}{6}
\biggl\{ 4(g^{\mu \sigma }g^{\lambda \nu }+g^{\sigma \nu }g^{\mu \lambda }
-g^{\sigma \lambda }g^{\mu \nu }) \biggr\}
\epsilon ^{r}_{\mu }\epsilon ^{s}_{\nu }\epsilon ^{s'}_{\lambda }\epsilon ^{r'}_{\sigma } .
 \eqno{(5.1.c)} $$
Thus, the total amplitude can be calculated as
$$ \hspace*{-6cm} \frac{1}{2}M = M_{a}+M_{b}+M_{c} 
= -\frac{(ie)^4}{(2\pi)^4}i\pi^2 \frac{1}{6} \ 4 \ \biggl\{
g^{\mu \sigma }g^{\lambda \nu }+g^{\mu \nu }g^{\sigma \lambda }
-g^{\mu \lambda }g^{\sigma \nu } $$
$$ \hspace*{.5cm}
g^{\mu \lambda }g^{\sigma \nu }+g^{\mu \nu }g^{\sigma \lambda }
-g^{\mu \sigma }g^{\lambda \nu }
$$
$$ \hspace*{2.2cm}
g^{\mu \sigma }g^{\lambda \nu }+g^{\sigma \nu }g^{\mu \lambda }
-g^{\sigma \lambda }g^{\mu \nu }
\biggr\}
\epsilon ^{r}_{\mu }\epsilon ^{s}_{\nu }\epsilon ^{s'}_{\lambda }\epsilon ^{r'}_{\sigma } 
$$
$$ \hspace*{-.8cm}
= -\frac{(ie)^4}{(2\pi)^4}i\pi^2\frac{1}{6} \ 4 \ \Bigl(
g^{\mu \sigma }g^{\lambda \nu }+g^{\mu \nu }g^{\sigma \lambda }
+g^{\mu \lambda }g^{\sigma \nu } \Bigr)
\epsilon ^{r}_{\mu }\epsilon ^{s}_{\nu }\epsilon ^{s'}_{\lambda }\epsilon ^{r'}_{\sigma }
$$
Therefore, we can write $M$ as
$$ M = -i \ \frac{4}{3} \ \alpha^2 \Bigl(
g^{\mu \sigma }g^{\lambda \nu }+g^{\mu \nu }g^{\sigma \lambda }
+g^{\mu \lambda }g^{\sigma \nu } \Bigr) 
\epsilon ^{r}_{\mu }\epsilon ^{s}_{\nu }\epsilon ^{s'}_{\lambda }\epsilon ^{r'}_{\sigma }
 \eqno{(5.2)} $$ 
where $ \alpha = \frac{e^2}{4\pi} $.


\section{Photon-Photon Scattering}
Now, the photon-photon scattering cross section can be written as \cite{LL}
$$ \frac{d\sigma }{d\Omega }
 = \frac{1}{64\pi^2} \frac{1}{(2\omega )^2} |M|^2
 \eqno{(6.1)} $$
where we sum up the final states of the photon polarization state and 
make average of the initial polarization states. 
$$ \hspace*{-5.2cm} |M|^2
 = \ \frac{1}{4} \sum_{rr'ss'}|M^{rr'ss'}|^2 . $$
Among the above, the nonvanishing term is
$$ = \ \frac{1}{4}\Bigl\{
 |M^{1111}|^2+|M^{2222}|^2+|M^{1122}|^2+|M^{2211}|^2
$$
$$ \hspace*{1cm}
+|M^{1221}|^2+|M^{2112}|^2+|M^{1212}|^2+|M^{2121}|^2
 \Bigr\} $$


Since we know that 
$$ M^{1122} = M^{2211} \hspace*{2cm}  M^{1221} = M^{2112}
 \hspace*{2cm} M^{1212} = M^{2121} 
$$
we obtain
$$
 \frac{1}{4} \sum_{rr'ss'}|M^{rr'ss'}|^2
 = \frac{1}{4} \ \Bigl\{
|M^{1111}|^2+|M^{2222}|^2+2|M^{1122}|^2+2|M^{1221}|^2+2|M^{1212}|^2
\Bigr\} .
$$
Further, we find
$$ \hspace*{-1.5cm} M^{1111} = -i\frac{4}{3} \alpha^2 \cdot 3 $$
$$ M^{2222} = -i\frac{4}{3} \alpha^2 (1+2\cos^2\theta ) $$
$$ \hspace*{-1.1cm} M^{1122} = -i\frac{4}{3} \alpha^2 \cos \theta  $$
$$ \hspace*{-1.4cm} M^{1221} = i\frac{4}{3} \alpha^2 \cos \theta $$
$$ \hspace*{-1.8cm} M^{1212} = i\frac{4}{3} \alpha^2 \cdot 1 $$
and thus、
$$ \frac{1}{4} \sum |M|^2
= \frac{16}{9} \alpha^2 \ \bigl\{ 3+2 \cos^2 \theta +\cos^4 \theta \bigr\} . $$
Therefore, for $\frac{\omega}{m} \ll 1$, we find
$$ \frac{d\sigma }{d\Omega }
= \frac{1}{(6\pi)^2} \ \frac{\alpha ^4}{(2\omega)^2} \
 \bigl( 3+2 \cos^2 \theta +\cos^4 \theta \bigr)
 \eqno{(6.2)} $$
where $\alpha \equiv \frac{e^2}{4\pi}$ \vspace*{0.5\baselineskip}. 
This is the scattering cross section in $\frac{\omega}{m} \ll 1$. 
This result is very different from the one obtained by 
Euler-Heisenberg \cite{HE},\cite{LL}. 
The calculated result of Euler-Heisenberg can be written as 
$$ \frac{d\sigma}{d\Omega} 
= \frac{139\alpha^4}{(180\pi)^2 m^2} \ \Bigl( \frac{\omega}{m} \Bigr)^6
 \ (3+\cos^2 \theta )^2 . \eqno{(6.3)} $$
In particular, the energy dependence of the cross section is completely 
different from each other. At the low energy limit, the new cross section 
becomes larger while the Euler-Heisenberg cross section becomes smaller. 
At $\omega \simeq 1 \ {\rm eV}$, the Euler-Heisenberg cross section
 with 90 degree becomes
$$ \frac{d\sigma}{d\Omega} = \ \frac{139 \alpha^4}{(60\pi)^2 m^2}
\biggl( \frac{\omega}{m} \biggr) ^6 \ \simeq \ 9.3 \ \times \ 10^{-67} \ cm^2 \
 \simeq \ 9.3 \ \times \ 10^{-43} \ b $$
which is extremely small. This suggests that the Euler-Heisenberg cross section is
 practically impossible to measure since it is too small. 
On the other hand, the present calculation predicts the cross section at 
$\omega \simeq 1$ eV 
$$ \frac{d\sigma}{d\Omega} = \ \frac{3 \alpha^4}{(12\pi)^2\omega^2}
 \ \simeq \ 2.3 \ \times \ 10^{-21} \ cm^2 \
 \simeq \ 2.3 \ \times \ 10^3 \ b . $$
This is rather large, and as far as the magnitude of the cross section is 
concerned, it should be indeed measurable. 
These problems are discussed in detail in \cite{FK}.


\section{Why  is  Euler-Heisenberg  Lagrangian incorrect ?}
Here, we briefly discribe the physical reason why the Euler-Heisenberg is
 incorrect.\\
\subsection{Euler-Heisenberg Lagrangian density}
The Euler-Heisenberg Lagrangian density is given as
$$ \mathcal{L} = \frac{2\alpha^2}{45m^4}
 \Bigl[(\bm{E}^2-\bm{B}^2)^2+7(\bm{E}\cdot \bm{B})^2 \Bigr] . \eqno{(7.1.a)} $$ 
They first calculate the vaccum polarization effects
 when they apply the electromagnetic fields to the Dirac vaccum states
 which are composed out of negative energy fermions. In this case,
 they obtained the effects of the vaccum polarization, and constructed
 the effective Lagrangian density which can simulate the results.
 The basic problem is that they treated the vacuum state as if it were
 dielectric matter.
\subsection{Karplus and Neuman's calculation}
Now, we discuss the result calculated by Karplus and Neuman
 who started their calculation from the QED Lagrangian density. Here, we want to clarify
 why Karplus and Neuman obtained the same results as that of Euler-Heisenberg,
 in spite of the fact that Karplus and Neuman employed the modern field theory
 terminology. First, we rewrite the Euler-Heisenberg Lagrangian density as
$$ \mathcal{L} = -\frac{\alpha^2}{180m^4} \Bigl\{
 5(F_{\mu \nu}(x)F^{\mu \nu}(x))^2
-14F_{\mu \nu }(x)F^{\nu \lambda }(x)F_{\lambda \sigma }(x)F^{\sigma \mu }(x)
 \Bigl\} \eqno{(7.1.b)} $$
where
$$ F^{\mu \nu }(x) = \partial ^{\mu }A^{\nu }(x)-\partial ^{\nu }A^{\mu }(x) . $$
Karplus and Neuman calculated first the QED evaluation based on
 the QED Lagrangian density. However, they always imposed the conditions
 that they should be able to reproduce the results of
 the Euler-Heisenberg calculation. This must be a mistery why they believed
 that they should get the same result as the one by Euler-Heisenberg.
 However, this may well be conected to the additional conditions
 which are often called "gauge condition" even though there is
 no physical reason for this gauge condition \cite{FK2}.



\section{conclusion}
We have presented the new QED calculation of photon-photon scattering at low energy. 
The calculation is straightforward since there is no logarithmic divergence
in the evaluation of the photon-photon scattering Feynman diagrams.
The result is very different from the Euler-Heisenberg calculation, and therefore
the photon-photon cross section should be measured by experiments in order to clarify
which of the cross section should be preferred by nature.

\section{acknowledgements}

The author is grateful to Prof. T. Fujita for his helpful discussions and comments.




\begin{thebibliography}{99}

\bibitem{EK}
H. Euler and B. Kockel, Naturwissensch. {\bf 23}, 246 (1935)


\bibitem{Euler}
H. Euler, Ann. der. Physik. {\bf 26}, 398 (1936)


\bibitem{HE}
W. Heisenberg and H. Euler, Z.Phys. {\bf 98}, 714 (1936)


\bibitem{LL}
V.B. Berestetskii, E.M. Lifshitz and L.P. Pitaevskii,
 "Relativistic Quantum Theory", (Pergamon Press, 1974)


\bibitem{KN}
R. Karplus and M. Neuman, Phys. Rev. {\bf 80}, 380 (1950)


\bibitem{KN2}
R. Karplus and M. Neuman, Phys. Rev. {\bf 83}, 776 (1951)


\bibitem{nishi} 
K. Nishijima, ``Fields and Particles", (W.A. Benjamin, INC, 1969)


\bibitem{mandl}
F. Mandl and G. Shaw, "Quantum Field Theory", 
(John Wiley \& Sons, 1993)


\bibitem{FK}
T. Fujita and N. Kanda, A Proposal to Measure Photon-Photon Scattering
 (to be published)
 
 
\bibitem{FK2}
T. Fujita and N. Kanda, "Tomonaga's Conjecture on Photon Self-Energy",
 physics.gen-ph/1102.2974 \\ 











\end{thebibliography}
\end{document}